\documentclass[type=submission,name=CES]{oae}

\usepackage{siunitx}
\usepackage{booktabs}
\usepackage{mathtools}

\documentsetup{
	title = {Stability-Preserving Automatic Tuning of PID Control \\with Reinforcement Learning},
	authors = {
		author1 = {
			name         = {Ayub I. Lakhani},
			affiliations = {affil1},
		},
		author2 = {
			name         = {Myisha A. Chowdhury},
			affiliations = {affil1},
		},
		author3 = {
			name         = {Qiugang Lu},
			affiliations = {affil1},
		},
	},
	affiliations = {
		affil1 = {
			Department of Chemical Engineering, Texas Tech University, Lubbock, TX 79409, USA
		},
	},
	correspondence-to = {
		Prof. Qiugang Lu, Department of Chemical Engineering, Texas Tech University, P.O. Box 43121, Lubbock, TX 79409-3121, USA. E-mail: jay.lu@ttu.edu; ORCID: 0000-0002-1478-3785    
	},
	short-author      = {Surname \textit{et al}.},
	journal           = {Complex Engineering Systems},
	short-journal     = {Complex Eng Syst},
	how-to-cite       = {},
	received          = {date month year},  
	first-decision    = {},
	revised           = {},
	accepted          = {},
	published         = {},
	academic-editor   = {},
	copy-editor       = {},
	production-editor = {},
	doi               = {10.20517/ces.xxxx.xx},
	year              = {Year},
	volume            = {Volume},
	end-page          = {Number},
}

\usepackage{hyperref}

\begin{document}
	
	\maketitle
	
	\begin{abstract}
		PID control has been the dominant control strategy in the process industry due to its simplicity in design and effectiveness in controlling a wide range of processes. However, most traditional PID tuning methods rely on trial and error for complex processes where insights about the system is limited and may not yield the optimal PID parameters. To address the issue, this work proposes an automatic PID tuning framework based on reinforcement learning (RL), particularly the deterministic policy gradient (DPG) method. Different from existing studies on using RL for PID tuning, in this work, we explicitly consider the closed-loop stability throughout the RL-based tuning process. In particular, we propose a novel episodic tuning framework that allows for an episodic closed-loop operation under selected PID parameters where the actor and critic networks are updated once at the end of each episode. To ensure the closed-loop stability during the tuning, we initialize the training with a conservative but stable baseline PID controller and the resultant reward is used as a benchmark score. A supervisor mechanism is used to monitor the running reward (e.g., tracking error) at each step in the episode. As soon as the running reward exceeds the benchmark score, the underlying controller is replaced by the baseline controller as an early correction to prevent instability. Moreover, we use layer normalization to standardize the input to each layer in actor and critic networks to overcome the issue of policy saturation at action bounds, to ensure the convergence to the optimum. The developed methods are validated through setpoint tracking experiments on a second-order plus dead-time system. Simulation results show that with our scheme, the closed-loop stability can be maintained throughout RL explorations and the explored PID parameters by the RL agent converge quickly to the optimum. Moreover, through simulation verification, the developed RL-based PID tuning method can adapt the PID parameters to changes in the process model automatically without requiring any knowledge about the underlying operating condition, in contrast to other adaptive methods such as the gain scheduling control.

		\paragraph{Keywords:}Reinforcement learning, PID tuning, Closed-loop stability, Deterministic policy gradient

	\end{abstract}

	\section{1. Introduction}
	
	Proportional-Integral-Derivative (PID) controllers have been the dominant type of controller for the process industry accounting for more than 80\% of industrial process control \cite{boubertakh2010tuning}. Such a popularity arises from their simplicity in the structure for design and effectiveness for controlling practical systems \cite{ziegler1942optimum,borase2020review}. However, some roadblocks still exist for PID design to enable a higher level of automation and adaptivity for controlling complex processes. First, the control performance of PID strictly relies on the setting of its parameters. Despite the fact that some empirical guidelines are available for tuning PID control even for complex systems, the resultant PID parameters may not be the optimal values \cite{lee2020reinforcement}. As reported in \cite{zhong2019toward}, most conventional methods use trial and error to tune PID for complex systems where insights about the system may be limited . Therefore, how to automate the search of optimal PID parameters becomes an interesting topic \cite{guan2021design}. Second, the parameters of traditional PID controllers generally remain fixed after tuning. As a result, they are mainly applicable for controlling time-invariant systems. For time-varying systems, as frequently encountered in practice such as robotics and vehicles, the lack of adaptivity of traditional fixed-parameter PID controllers makes them unable to maintain high control performance under such scenarios \cite{teoh1991implementation}. 
	
Traditional PID tuning methods can be roughly classified into three categories: heuristic tuning, rule-based tuning, and optimization-based (or model-based) tuning \cite{bucz2018advanced}. Heuristic tuning often relies extensively on trial and error based on the understanding of the role of each PID parameter. This method is easy to implement, however, it can be time-consuming and cannot guarantee to reach a robust or optimal solution \cite{bucz2018advanced}. Rule-based tuning establishes simple models (often first-order plus dead-time model) to approximate the process based on the step test, including methods such as Ziegler-Nichols,  Cohen-Coon, Kappa-Tau, and Lambda tuning \cite{seborg2016process}. These methods are widely used, however, they are sensitive to the discrepancies between the true process and the approximation model. Optimization-based methods can find the optimal PID parameters given the availability of an accurate process model along with the desirable engineering specification. However, a sufficiently accurate model is required for such methods \cite{abushawishpid}, which can be difficult in practice.
	
	Adaptive PID has been proposed in the literature to mitigate the above non-adaptivity issue. These control strategies can be classified into model-based approaches \cite{chang2003multivariable,yu2007stable}, evolution optimization-based approaches \cite{zhou2005optimal}, and neural network-based approaches \cite{chen2004applying}. Model-based approaches assume the presence of an accurate model that can represent the true dynamics exactly to allow for the adaptivity of PID controllers, which does not hold in practice due to the difficulty in accessing such high-quality models for complex dynamical systems \cite{hou2016overview}. Evolution optimization-based approach is hard to reach real-time and online adaptation due to the slow computation speed \cite{wang2007proposal}. Adaptive PID based on neural networks employs supervised learning to optimize network parameters for enabling adaptivity. This is limited by the fact that the teaching signal is hard to acquire \cite{guan2021design}. To this end, more advanced machine learning techniques have been adopted to promote the user-free automatic tuning and adaptivity of traditional PID controllers. 
	
	Among emerging machine learning techniques for further empowering PID control, reinforcement learning (RL) has shown unique advantages to address the above issues. As a sequential decision-making method, RL can iteratively optimize an objective, usually black-box, via guided trial-and-error exploration \cite{sutton2018reinforcement}. First, due to the data-driven nature, automatic tuning of PID parameters can be achieved where expert knowledge is not needed. Second, RL can learn the optimal strategy in real-time and online by interacting with the environment. PID tuning is essentially a black-box optimization problem where the relation between tuning parameters and control performance is unknown. Optimal PID parameters shall be the ones that optimize the control performance. In light of this observation, RL has been preliminarily employed to facilitate the PID tuning \cite{shin2019reinforcement}. Typical model-free RL algorithms can be broadly grouped into value-based methods (e.g., Q-learning \cite{watkins1992q}, SARSA \cite{rummery1994line}, and temporal difference (TD) learning \cite{sutton2018reinforcement}), and policy-based algorithms (e.g., policy gradient \cite{sutton2000policy}). Value-based methods learn an action or state value function, from which the optimal policy can be derived. In contrast, policy-based methods directly learn the optimal policy from the experience, therefore, they have better convergence and are effective for continuous action space. In particular, the deterministic policy gradient (DPG) method, proposed by DeepMind \cite{silver2014deterministic}, receives wide attention due to its sample efficiency by considering deterministic rather than stochastic policies. As a result, the usage of off-policy actor-critic-based DPG algorithm has shown attractive prospects for addressing the PID tuning problem \cite{lawrence2020reinforcement}.
	
	Treating the PID tuning as an RL problem has been reported in the literature where different RL algorithms have been employed \cite{lee2020reinforcement,qin2018improve}. Actor-critic learning based on a single RBF network is proposed in \cite{wang2007proposal} to enable an adaptive self-tuning PID control, which is later implemented to wind turbine control \cite{sedighizadeh2008adaptive}. The combination of fuzzy PID and RL is reported in \cite{boubertakh2006optimization,boubertakh2010tuning}. Applications of RL-based self-tuning PID include soccer robot \cite{el2013application}, multicopter \cite{park2019multicopter}, and  human-in-the-loop physical assistive control \cite{zhong2019toward}. For process control, reports in \cite{lawrence2020reinforcement,lawrence2020optimal} consider the PID tuning in sample-by-sample and episodic modes, respectively. A sample-efficient deep RL with episodic policy transfer is proposed for PID tuning in a robotic catheter system in \cite{omisore2021novel}. Despite these rapid advancements, maintaining the closed-loop stability during RL search has not been thoroughly studied. Note that the ultimate goal of RL-based PID tuning is to accomplish online and adaptive tuning to maintain high performance even in the presence of changes in operating conditions. Such expectations require that the attempted PID parameters be located in the stable region to avoid unstable response of the closed-loop process.  
	
	The main contribution of this work is to develop a stability-preserving automatic PID tuning approach based on RL, particularly the DPG algorithm. The majority of the literature in this direction so far has focused on establishing an adaptive auto-tuning framework with RL approaches. However, how to maintain the closed-loop stability during the exploration of the PID parameter space by the RL agent has not been well studied. For instance, as one of the few reports that briefly discuss the stability issue, the authors in \cite{lawrence2020optimal} use an anti-windup strategy to force the input to be within the bound so as to maintain the boundedness of the output in the presence of unstable PID controller during RL search. However, this paper does not directly tackle the stability issue in a rigorous way. In contrast to existing work, our approach explicitly accounts for the closed-loop stability issue during RL exploration. For the proposed framework, we use a conservative but stable PID controller as a baseline. A supervisor mechanism monitors the closed-loop behavior of process variables at every time step throughout a training episode (i.e., one complete step test). Once the running reward (e.g., cumulative tracking error) exceeds a threshold, the employed PID controller will be replaced by the baseline controller to correct the process variables and prevent instability. In this work, we primarily focus on the automatic tuning aspect of RL-based PID tuning with stability preservation. In the simulation, we demonstrate the adaptivity of our method with respect to the changes in the system model throughout the offline episodic tuning. However, how to fully accomplish the online adaptivity to varying operating conditions with stability preservation will be tackled in our future work. 
	
	This paper is outlined as follows. In Section 2, we  introduce the fundamentals about the form of PID control used in this work and the DPG algorithm that will be adapted in our proposed method. Section 3 is devoted to the introduction of the proposed episodic RL-based PID auto-tuning method, where the closed-loop stability can be maintained throughout the RL training. In Section 4, simulation examples are provided to demonstrate the effectiveness and adaptivity of the proposed techniques. This paper concludes in Section 5. 
	

	\section{2. Preliminaries}\label{Section: II}
	\subsection{PID Control}
	We consider the standard position form of PID control \cite{seborg2016process}:
	\begin{equation}
		u(t) = K_{p}\left[e(t) + \frac{1}{\tau_{I}} \int_{0}^{t} e(t) dt + \tau_{D} \frac{de(t)}{dt}\right], \label{eq: PID}
	\end{equation}
	where $e(t)=r^{sp}(t) - y(t)$ represents the tracking error between setpoint $r^{sp}(t)$ and controlled variable (CV) $y(t)$ at time $t$, and $u(t)$ is the manipulated variable (MV). The digital version to implement \eqref{eq: PID} is shown to be:
	\begin{equation}
		u(t_{n}) = K_{p}\left[e(t_{n}) + \frac{1}{\tau_{I}} I(t_n) + \tau_{D} D(t_n)\right], \label{eq: PID_discrete}
	\end{equation}
	where $I(t_n)=\sum_{i=1}^{n}e(t_i)$ and $D(t_n)=\frac{e(t_n)-e(t_{n-1})}{\Delta t}$ are, respectively, the integrated and differentiated errors. In this work, we consider the following actuator saturation
	\begin{equation}
		sat(u)=
		\begin{dcases}
			u_{min}, \quad & \text{if} \quad  u < u_{min}, \\
			u, \quad & \text{if} \quad  u_{min} \le u \le u_{max}, \\
			u_{max} \quad & \text{if} \quad  u > u_{max}. 
		\end{dcases} \label{eq: Saturation}
	\end{equation}
	We employ the \textit{anti-windup compensation} \cite{astrom1989integrator} to address the issue that the integral action continues to increase or decrease if the saturation persists. The presence of windup of the integral action can trigger the nonlinearity of the controller that may lead to instability of the closed-loop system \cite{lawrence2020reinforcement}. With the anti-windup compensation, the integral term does not accumulate once the control action reaches the upper or lower bound. Further, to avoid the sudden jump of the derivative of the error when the setpoint is adjusted, the \textit{derivative kick} will be overcome by using the differentiated output $\frac{y(t_{n-1})-y(t_n)}{\Delta t}$ to replace the differentiated error $\frac{e(t_n)-e(t_{n-1})}{\Delta t}$. 
	
	\subsection{Reinforcement Learning}
	
	\subsubsection{Introduction}
	
\begin{figure} [tb]
		\centering
		\includegraphics[width=2.5in]{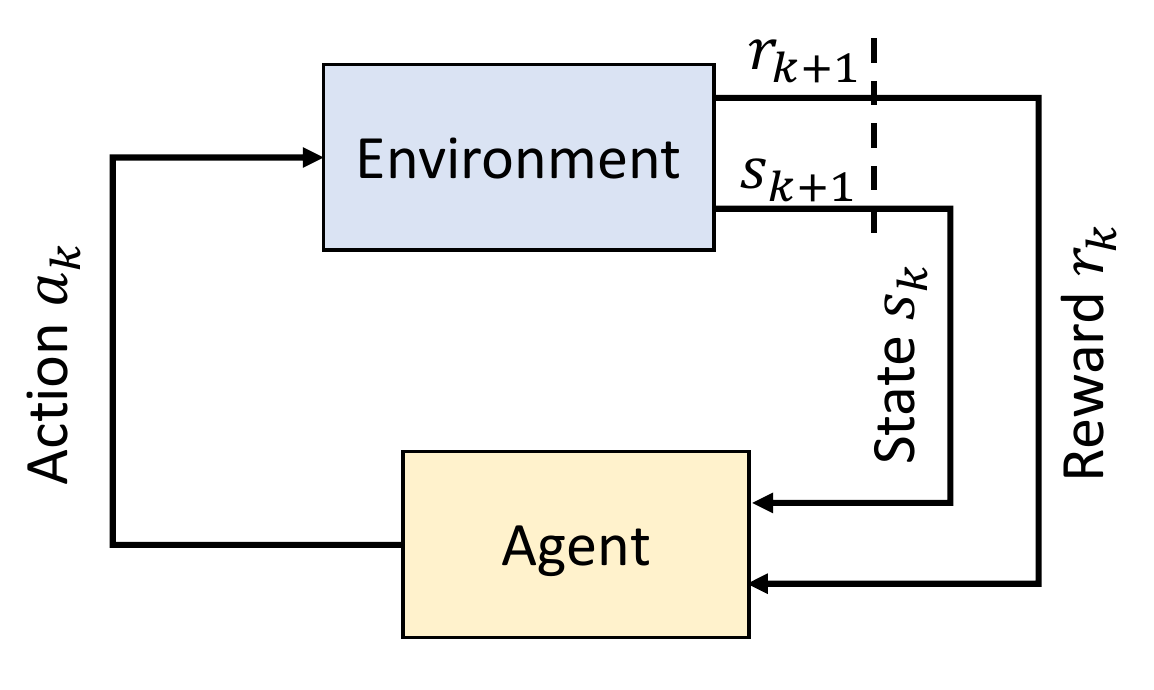}
		\caption{Illustration of the interaction between RL agent and environment.}
		\label{fig: Agent_environment}
	\end{figure}

	The standard reinforcement learning contains a learning agent that interacts with an environment. The agent (e.g., a controller) represents a decision-making mechanism, whereas the environment often represents an object (e.g., a plant or a process) \cite{spielberg2020deep}. The objective of an agent is to find the best policy such that the long-term cumulative reward (known as \textit{return}) is optimized by interacting with the environment that is often characterized by a Markov decision process (MDP) \cite{sutton2018reinforcement}. Specifically,  at each time step $k$, the agent observes a state $s_k \in\mathbb{S}$ of the environment and then generates a control action $a_k\in \mathbb{A}$ based on the policy. The symbols $\mathbb{S}$ and $\mathbb{A}$ denote the state set and action set, respectively. The control action is then deployed to the environment, where a reward value $r_{k+1}$ is assigned to the action at the given state as a performance measure. At the next time step, the state of the environment transits to $s_{k+1}$ which is determined by the previous control action $a_k$ and the internal dynamic behavior of the environment. After receiving the updated reward $r_{k+1}$ and state $s_{k+1}$, the agent adjusts its policy to generate the next control action $a_{k+1}$. Such iterations continue indefinitely for a continuous task or terminate after certain steps for an episodic task. Over iterations, the policy improves to generate better control actions that can give higher returns. Fig. \ref{fig: Agent_environment} illustrates the interaction between the RL agent and the environment. 
	
	The agent's policy is often described by a mapping $\pi: \mathbb{S} \to \mathcal{P}(\mathbb{A})$ where $\mathcal{P}(\mathbb{A})$ is the probability measure on set $\mathbb{A}$. The return starting from time step $k$ onward is defined as:
	\begin{equation}
		R_k = \sum_{i=k}^{\infty} \gamma^{i-k}r_{i}(s_i,a_i),
	\end{equation}
	where $\gamma \in [0,1]$ is the discounting factor that puts less weight to the reward far in the future to account for the uncertainty. For episodic tasks the sum above will stop at the specified terminal state. 
	
	The state value function $V^{\pi}: \mathbb{S}\to \mathbb{R}$ defines the \textit{expected} return at a given state if the policy follows $\pi$, i.e., 
	\begin{equation}
		V^{\pi}(s_k)=\mathbb{E}_{\pi}\left[R_k|s = s_k \right].
	\end{equation}
	Similarly, the action-value function $Q^{\pi}: \mathbb{S} \times \mathbb{A} \to \mathbb{R}$ represents the expected return if one starts with $s_k$ and takes action $a_k$, according to the policy $\pi$, expressed as
	\begin{equation}
		Q^{\pi}(s_k, a_k)=\mathbb{E}_{\pi}\left[R_k|s = s_k, a = a_k \right].
	\end{equation}
	With the Markov property, the recursive relation (Bellman expectation equation) between the $Q$ values at two consecutive time steps is shown to be:
	\begin{equation}
		Q^{\pi}(s_k, a_k)=\mathbb{E}_{s_{k+1}\sim p(\cdot|s_k,a_k)}\left[ r_{k}(s_k,a_k)  + \gamma \mathbb{E}_{a_{k+1}\sim \pi(\cdot|s_{k+1})}\left(Q^{\pi}(s_{k+1}, a_{k+1})\right)\right],
	\end{equation}
	where $p(\cdot|s_k,a_k)$ is the transition probability between consecutive states of the environment given $s_k$ and $a_k$, and $\pi(\cdot|s_k)$ is the control policy. The RL problem can be formulated as finding the optimal policy $\pi^{*}$ such that the associated state and action value functions outperform those from the other policies across all states and actions. Under the optimal policy, the Bellman optimality equation is formulated as
	\begin{equation}
		Q^{*}(s_k, a_k)=\mathbb{E}_{s_{k+1}\sim p(\cdot|s_k,a_k)}\left[ r_{k}(s_k,a_k)  + \gamma \max_{a^{\prime}\in\mathbb{A}}Q^{*}(s_{k+1}, a^{\prime})\right]. \label{eq: BellmanOptimality}
	\end{equation}
	The advantage of using $Q$ value functions is that once $Q^{*}$ is obtained, the corresponding optimal policy $\pi^{*}$ can be easily recovered as 
	\begin{equation}
		\pi^{*}(a|s_k)=\arg \max_{a\in\mathbb{A}} Q^{*}(s_k, a).
	\end{equation}
	Value-based methods have been well developed to solve the Bellman expectation or optimality equations. Among these methods, Q-learning has been the most widely used due to its off-policy nature and ability to discover the global optimum \cite{watkins1989learning,watkins1992q}. Q-learning solves the Bellman optimality equation \eqref{eq: BellmanOptimality} via temporal difference (TD):
	\begin{equation}
		Q(s_k,a_k) \leftarrow Q(s_k,a_k) + \alpha \delta, \label{eq:Q-learning}
	\end{equation}
	where $\delta=r(s_k,a_k) + \gamma \max_{a^{\prime}\in\mathbb{A}} Q(s_{k+1},a^{\prime})-Q(s_k,a_k)$ is the TD error. TD method is model-free (in contrast to model-based methods such as dynamic programming) and does not require the complete trajectory of an episode. Therefore, TD method is widely used for continuous tasks without a terminal state. Despite these advantages, value-based methods often assume that  states and actions are discrete and the value functions can be stored in a table to sweep through. However, many practical problems, including process control, have continuous states and actions. Therefore, policy-based methods and functional approximators (FAs) are developed to address these issues, as described below.
	
	To generalize Q-learning to continuous state and action spaces, neural networks have been extensively used as FAs to represent $Q$ values. In particular, a neural network $Q(s_k,a_k,w)$, parameterized by $w$, is often used to approximate $Q(s_k,a_k)$. In other words, the objective is to choose the optimal $w$ to minimize 
	\begin{equation}
		J(w) = \mathbb{E}_{\pi}[Q(s_k,a_k)-Q(s_k,a_k,w)]^2. \label{eq: FA}
	\end{equation}
	Note that the objective above is a supervised learning problem with $Q(s_k,a_k)$ being the target whose value is unknown during Q-learning. Therefore, similar to the TD method for Q-learning \eqref{eq:Q-learning}, the incurred reward and next-step parameterized $Q(s_{k+1},w)$ are often used together to approximate $Q(s_k,a_k)$, i.e., assuming that $r(s_k,a_k) + \gamma \max_{a^{\prime}\in\mathbb{A}} Q(s_{k+1},a^{\prime},w)\approx Q(s_k,a_k)$. With this approximation, stochastic gradient descent (SGD) can be used to solve \eqref{eq: FA} where the update of parameter $w$ is shown to be
	\begin{equation}
		w_{k+1} = w_k + \Delta w, \label{eq: FA_update}
	\end{equation}
	where $\Delta w = \alpha \left[r(s_k,a_k) + \gamma \max_{a^{\prime}\in\mathbb{A}} Q(s_{k+1},a^{\prime},w)-Q(s_k,a_k,w) \right]\nabla_{w}Q(s_k,a_k,w)$. The above combination of Q-learning and neural-network-based FAs is known as deep Q-network (DQN). 
	
	\subsubsection{Deterministic Policy Gradient}\label{sec: DPG}
	In contrast to value-based methods where action or state value functions must be obtained before acquiring the optimal policy, policy-based methods directly search for the best policy to optimize the RL objective:
	\begin{equation}
		J(\theta) = \mathbb{E}_{\pi_{\theta},s_k}\left[R_1|s_0\right], \label{eq: PG_obj}
	\end{equation}
	where $s_0$ is the starting state of the task and $\pi_{\theta}$ is a neural network, known as \textit{actor network}, parameterized by $\theta$, to approximate the continuous action space $\pi_{\theta}: \mathbb{S}\to\mathbb{A}$. The policy gradient algorithm uses SGD to update $\theta$ to maximize the objective in \eqref{eq: PG_obj}, such that 
	\begin{equation}
		\theta_{k+1}\leftarrow \theta_{k} + \alpha_k \nabla_{\theta} J(\theta)|_{\theta=\theta_k}, \label{eq: PG_update}
	\end{equation}
	where $\alpha_k$ is the learning rate, and the gradient of $J(\theta)$, using the policy-gradient theorem \cite{sutton2000policy}, turns out to be 
	\begin{equation}
		\nabla_{\theta} J(\theta) = \mathbb{E}_{\pi_{\theta},s_k}\left[ Q^{\pi}(s_k,a_k)\nabla_{\theta} \log{\pi_{\theta}(a_k|s_k)}\right], \label{eq: PG_gradient}
	\end{equation}
	which is in terms of the $Q$ value function. Note that in the implementation of the policy gradient above, the expectation is replaced by samples (i.e., using SGD) for updating the parameter $\theta$. Moreover, the $Q$ value in \eqref{eq: PG_gradient} is replaced by its neural network FA $Q^{\pi}(s_k,a_k,w)$ to accommodate continuous state and action spaces, known as the \textit{critic network}. The parameters of actor and critic networks can be updated sequentially using \eqref{eq: FA_update} and \eqref{eq: PG_update}. This forms the well-known \textit{actor-critic} architecture \cite{degris2012off}. The actor aims to find the optimal policy and the critic serves as a judge to assess the current policy prescribed by the actor. 
	
	Stochastic policies often demand a large number of samples for evaluating the policy gradient \eqref{eq: PG_gradient}, especially for high-dimensional action space. As an extension of DQN, \textit{deterministic policy gradient} (DPG) is proposed by Silver et al. \cite{silver2014deterministic} where the deterministic policy is given as $\mu_{\theta}: \mathbb{S}\to \mathbb{A}$. That is, for a given state $s_k$, the actor returns $a_k = \mu_{\theta}(s_k)$ with probability one. Then the gradient of $J(\theta)$ becomes
	\begin{equation}
		\nabla_{\theta} J(\theta) = \mathbb{E}_{s_k}\left[\nabla_{a} Q^{\mu}(s_k,a)|_{a=\mu_{\theta}(s_k)}\nabla_{\theta}\mu_{\theta}(s_k)\right]. \label{eq: DPG_gradient}
	\end{equation}
	Compared with the stochastic counterpart \eqref{eq: PG_gradient}, the gradient of the deterministic policy above has expectation with respect to only the state transition probability, thereby the required number of data samples can be significantly reduced.
	
	As an extension of DQN, the DPG algorithm inherits some techniques adopted by DQN. (i): As shown in \eqref{eq: BellmanOptimality}, the target of Q-learning problem with FA can be represented as $r(s_k,a_k) + \gamma \max_{a^{\prime}\in\mathbb{A}} Q(s_{k+1},a^{\prime},w)$, which depends on the parameter $w$ that is to be optimized. That is, the target is varying and this causes significant difficulty for the supervised learning. \textit{Fixed Q target} is one solution to alleviate this issue, in which a \textit{critic target network} $Q(s_{k+1},a^{\prime},w^{\prime})$ with the same architecture as the critic network is used but its parameter $w^{\prime}$ does not change rapidly; (ii): The target in \eqref{eq: BellmanOptimality} contains the maximization over $a^{\prime}\in\mathbb{A}$ that can be expensive since the $Q$ value function is a complex network with continuous inputs. To address this problem, a \textit{target actor network} $\mu_{\theta^{\prime}}(s_{k+1})$ is used to approximate the optimal policy that maximizes $Q(s_{k+1},a^{\prime},w^{\prime})$. That is, the target becomes $r(s_k,a_k) + Q(s_{k+1},\mu_{\theta^{\prime}}(s_{k+1}),w^{\prime})$. In summary, there are four networks involved in the DPG framework: (1) actor network $\mu_{\theta}(s_k)$ for selecting action based on the current state; (2) critic network $Q(s_k,a_k,w)$ for evaluating the current $Q$ value function; (3) actor target network $\mu_{\theta^{\prime}}(s_{k+1})$ for selecting the bast action for maximizing the next-step $Q$ value; and (4) critic target network $Q(s_{k+1},\mu_{\theta^{\prime}}(s_{k+1}),w^{\prime})$ for evaluating the optimal $Q$ value for next state. To alleviate the dynamic correlations between samples, the \textit{replay buffer} is often used to store a large number of past transition pairs $\left\{s_k,a_k,r_k,s_{k+1}\right\}$. When updating the parameters in DPG, a batch $B$ of  past experience from the replay buffer is randomly selected, and they are used to formulate two supervised learning problems for the critic and actor, respectively. The detailed DPG algorithm can be found in \cite{silver2014deterministic,lillicrap2015continuous}. 

	\section{3. Stability-Preserving PID Tuning with DPG}\label{Sec: III}
	
	In this section, we demonstrate the procedures of adapting the above DPG algorithm to the PID tuning problem with stability guarantee. 
	
	\begin{figure} [tb]
		\centering
		\includegraphics[width=5in]{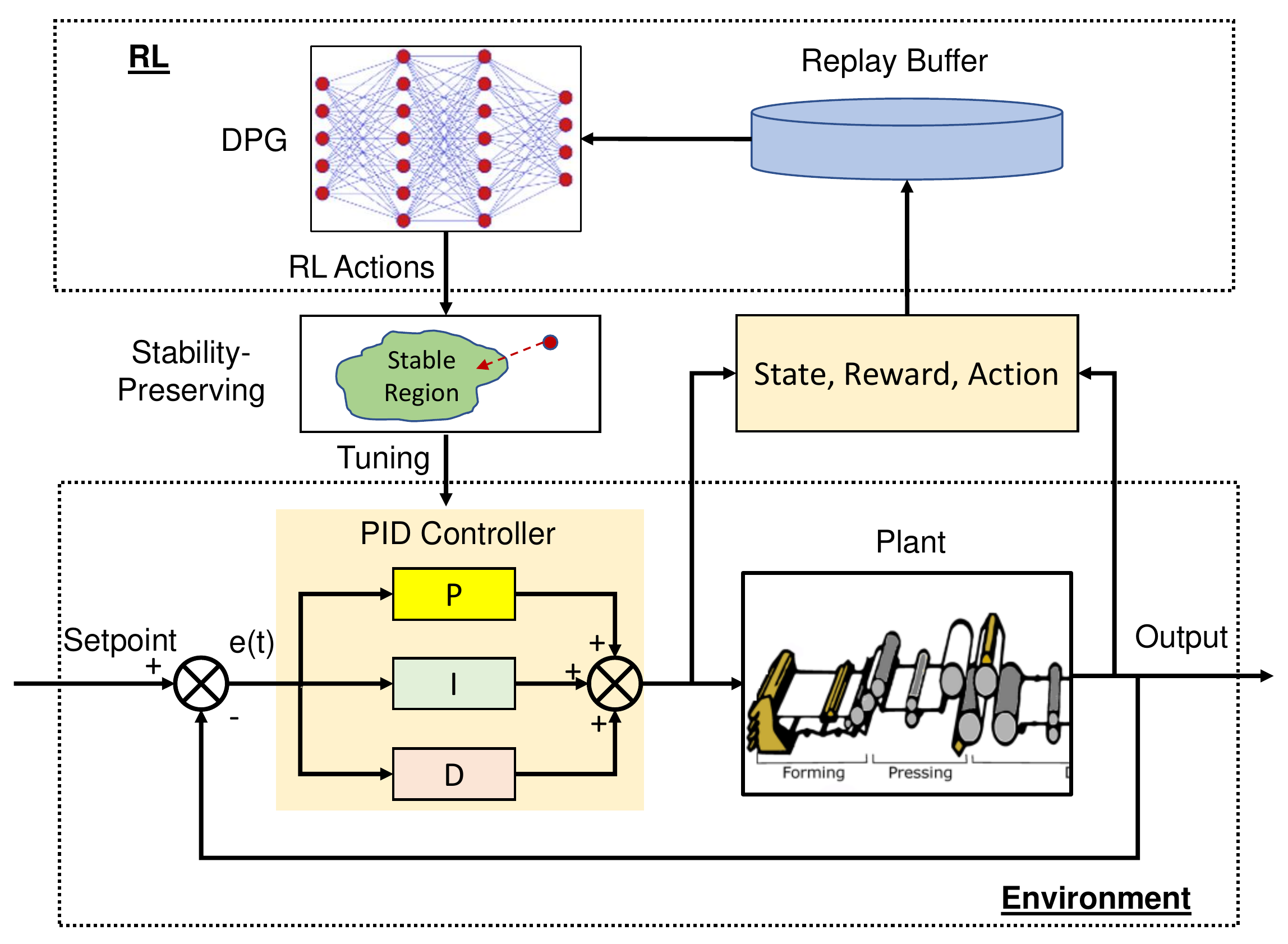}
		\caption{Illustration of the stability-preserving RL-based PID tuning.}
		\label{fig: PID_Tuning_Schematics}
	\end{figure}
	
	\subsection{PID Tuning with DPG}
	
	The overall structure of the proposed DPG-based PID tuning with stability guarantee is shown in Fig. \ref{fig: PID_Tuning_Schematics}. The proposed scheme is \textit{multi-scale} in time for RL training and closed-loop operation. For clarity, we define the \textit{update step} as the step that the parameters of RL agent are updated once, or equivalently, the parameters in the PID controller are changed once. We use $k\in \{0,1,\ldots,K\}$ to denote the $k$-th update step, where $K$ represents the maximum number of training episodes. In contrast, the \textit{operation step} is defined as one time step in a closed-loop operation, denoted as $t\in \{0,1,\ldots,T\}$, where $T$ is the length of the episode. Each closed-loop operation gives one sample of state and action transition of the environment. The parameters of the four networks in the DPG algorithm are updated once as soon as one episodic closed-loop operation is complete. Note that the PID parameters remain unchanged throughout one episode. One motivation for this setup is that by overseeing the entire closed-loop trajectories of process variables in an episode, the performance of the underlying PID parameters can be more comprehensively assessed by considering both point-based metrics such as overshoot, rise time, etc., and trajectory-based metrics such as integrated square error (ISE).  Moreover, as stated in \cite{lawrence2020reinforcement}, if the RL agent and PID parameters are updated at every operation step, as has been done by many works, the fast switching of controller parameters may lead to instability \cite{malmborg1996stabilizing}. The details of each block in Fig. \ref{fig: PID_Tuning_Schematics} are provided below.

	\subsubsection{Environment, State and Action} 
	In the proposed scheme, the closed-loop system with PID control is viewed as the environment. The outputs of the environment are the trajectories of process variables, such as MVs, CVs, and setpoint, throughout an episodic closed-loop operation. Specifically, after the $k$-th closed-loop operation, the trajectories of process variables are
	\begin{equation}
		\mathbf{y}_{k}=\left[y_0^{k},y_1^{k},\ldots,y_T^{k}\right]^\mathsf{T},~~ \mathbf{u}_{k}=\left[u_0^{k},u_1^{k},\ldots,u_T^{k}\right]^\mathsf{T}, \label{eq: Trajectories}
	\end{equation}
	where $\mathbf{y}_{k}$ and $\mathbf{u}_{k}$ are the CV and MV trajectories, respectively. Existing literature \cite{spielberg2020deep,shin2019reinforcement} generally stacks the entire input and output trajectories, possibly with setpoint and other information, into a tuple as the environment state. However, this can lead to the issue of overly high-dimensional input to the DPG networks, especially for slow processes that require a large episode length $T$ to reflect the overall response trend. As a result, the networks involved in DPG will be large in scale and this makes the training extremely challenging. In this work, to reduce the input dimension, we propose to use a subset of these trajectories as the environment state. To be specific, we select a subset of CVs and MVs at an equally-spaced interval over  $\mathbf{y}_{k}$ and $\mathbf{u}_{k}$. The environment state is defined as 
	\begin{equation}
		\mathbf{s}_k=\left[\tilde{\mathbf{y}}_k, \tilde{\mathbf{u}}_k, \tilde{\textbf{r}}^{sp}_k\right]^\mathsf{T},~~\tilde{\mathbf{y}}_k=[y_{0}^{k},y_{i}^{k},y_{2i}^{k},\ldots]^\mathsf{T}, ~~\tilde{\mathbf{u}}_k=[u_{0}^{k},u_{i}^{k},u_{2i}^{k},\ldots]^\mathsf{T}, ~~\tilde{\mathbf{r}}_k^{sp}=[r_{0}^{sp,k}, r_{i}^{sp,k}, r_{2i}^{sp,k},\ldots]^\mathsf{T}, \label{eq: Env_state}
	\end{equation}
	where $i$ represents the interval over which a subset of process variable is selected. In \eqref{eq: Env_state}, the setpoint information $\tilde{\textbf{r}}^{sp}_k$ is included to inform the RL agent of the tracking target information. This becomes important if an episode contains setpoint changes, a common scenario often met for online controller tuning with RL. In our work, for simplicity, we will only consider the situation where the setpoint is a constant. Scenarios with varying setpoints and online tuning will be tackled in our future work. The action $\mathbf{a}_k$ of the RL agent consists of the PID parameters: 
	\begin{equation}
		\mathbf{a}_k = \left[K_p, \tau_I, \tau_D\right].
	\end{equation}
	At the end of each episode, the RL agent receives  information from the environment, updates its parameters, and then delivers one set of PID parameters to the controller. The closed-loop system then operates under the new PID controller for the next episode.

	\subsubsection{Reward} 
	The proposed method will use the reward $r$ to assess the control performance of PID by monitoring closed-loop trajectories of CVs and MVs. Instead of relying on point-based metrics such as overshoot, settling time, etc., we will utilize trajectory-based metrics, particularly the ISE as the primary assessment method:
	\begin{equation}
		r = -\sum_{t=0}^{T} e_t^{2}, \label{eq: ISE}
	\end{equation}
	where $e_t = r_{t}^{sp}-y_{t}, t = 0,1,\ldots,T$, represents the tracking error. Note that more comprehensive assessment metrics consisting of both trajectories- and point-based metrics can easily be constructed as above. In addition, the amplitude of the MV can also be incorporated into the performance criterion with a weight reflecting its importance relative to the CV. In our scheme, the reward \eqref{eq: ISE} is computed using the entire closed-loop trajectories \eqref{eq: Trajectories}, rather than the subset of them selected for defining the state as in \eqref{eq: Env_state}. 
	
	\begin{figure} [tb]
		\centering
		\includegraphics[width=6.5in]{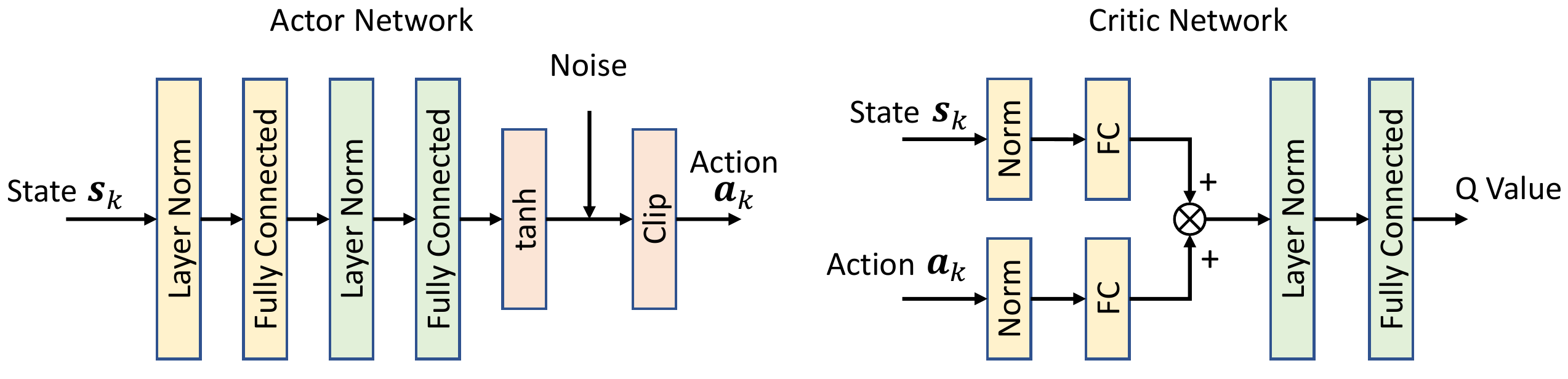}
		\caption{The structure of the actor and critic networks. Left: The actor network where layer normalization is used before each network layer. Decaying noise is added to the output to encourage exploration at the beginning of RL training. Right: The critic network that consumes state and action, and returns the $Q$ value for updates with policy gradient. }
		\label{fig: Actor_Critic_Network}
	\end{figure}

	\subsubsection{Agent} 
	As discussed in Section 2, our RL will utilize the DPG algorithm for training the agent. At the end of each closed-loop operation episode, the setpoint, MV, and CV trajectories will be consumed to extract the environment state, action (PID parameters used for this episode), and reward, as shown in Fig. \ref{fig: PID_Tuning_Schematics}. Such information is saved into the replay buffer. A batch of $B$ samples of past transition data $(s_k,a_k,r_k,s_{k+1})$ is then randomly sampled to be used as the dataset for updating network parameters in the agent. Fig. \ref{fig: Actor_Critic_Network} shows the structures of actor network $\mu_{\theta}(s_k)$, critic network $Q(s_k,a_k,w)$, and their target networks $\mu_{\theta^{\prime}}(s_{k+1})$, $Q(s_{k+1},\mu_{\theta^{\prime}}(s_{k+1}),w^{\prime})$ used by the agent in this work. During RL exploration, it is likely that the agent will attempt some PID parameters that are close to the unstable region of the parameter space. This could lead to closed-loop responses where CV and MV have overly large amplitudes, which has been observed in our simulation study. Without normalization, it is likely that such a large-amplitude input to the actor network will drive the output of the actor network to the saturation zone of the `tanh' function \cite{ioffe2015batch}. As a result, the returned action $\mathbf{a}_k$ always stays at the upper or lower bounds. Such a behavior prevents the agent from learning meaningful policies \cite{lillicrap2015continuous}. In this work, we use layer normalization \cite{ba2016layer} to mitigate this issue. For the actor network, in order to encourage the agent's exploration of PID parameter space at the beginning, we add a Gaussian noise after the `tanh' function. However, the variance of the noise decays over episodes so that the agent will focus more on exploitation at the later stage of the training. Note that the architectures (e.g., the number of layers and nodes for each layer) of these networks in Fig. \ref{fig: Actor_Critic_Network} for specific problems need fine tuning to optimize the performance of the trained RL agent. As shown in Fig. \ref{fig: PID_Tuning_Schematics}, the output (action) of the RL agent is the next set of PID parameters to be deployed to the closed-loop system. However, some PID parameters, if implemented, may give rise to unstable closed-loop responses that can cause disastrous consequences in practice. To preserve the stability and thus ensure safety, a novel mechanism is proposed to correct those poorly selected PID parameters before causing instability, as discussed in the next subsection.

	\begin{figure} [tb]
		\centering
		\includegraphics[width=4in]{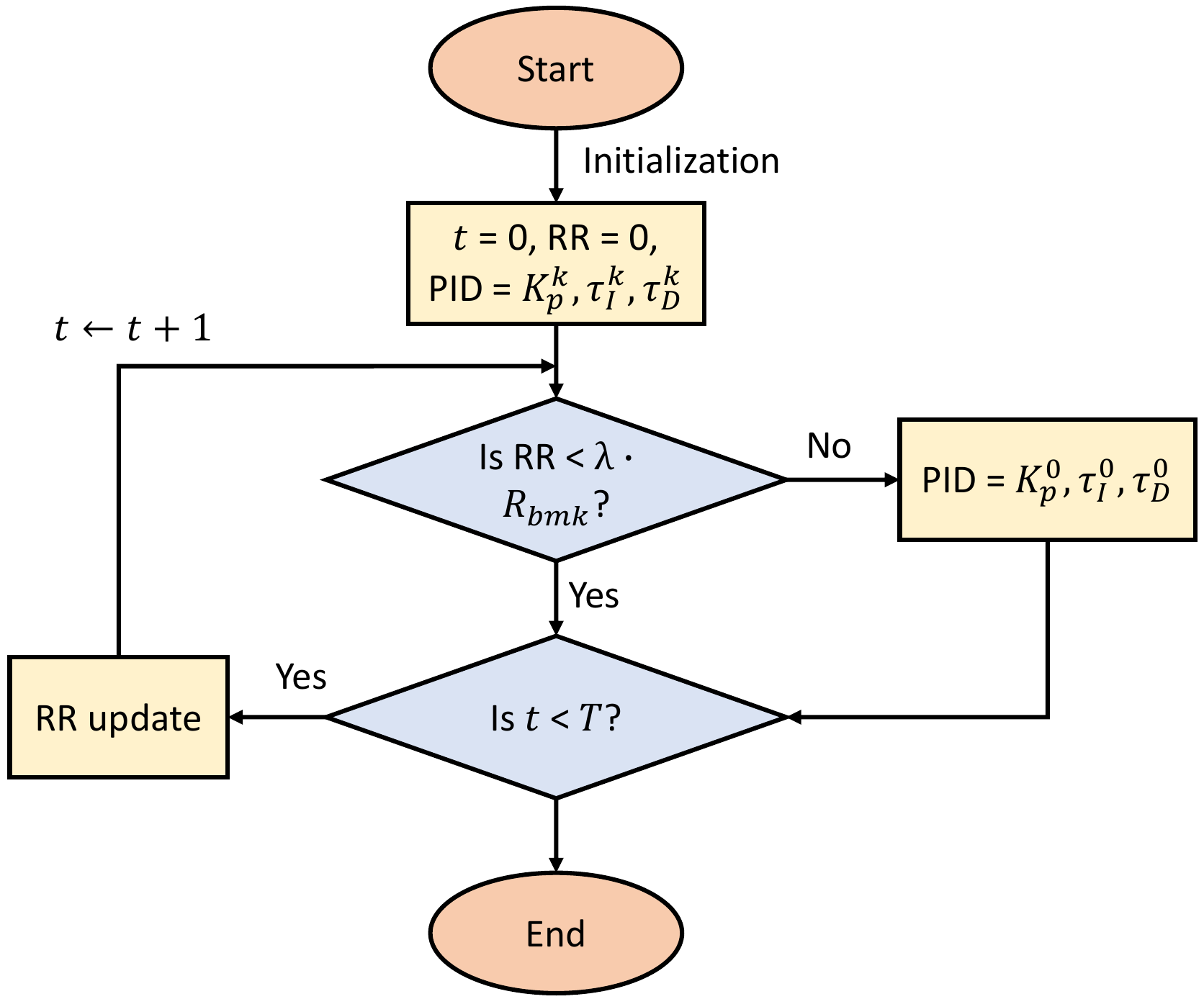}
		\caption{The flow chart for the proposed stability-preserving RL-based PID tuning approach. The running reward is denoted as ``RR''.}
		\label{fig: Flow_Chart}
	\end{figure}
	
	\subsection{Stability-preserving PID tuning with Baseline Controller}
	
	Ensuring the stability of the closed-loop system during RL exploration of PID parameters is critical for practical implementation. However, algorithms such as DPG employed by RL agent focus solely on maximizing the return \eqref{eq: PG_obj} that does not directly relate to the closed-loop stability, especially when the training episode has a short duration where the unstable behavior cannot be clearly manifested. Moreover, constrained RL algorithms \cite{liu2021policy} that restrict the searchable action space to ensure safety may not be applicable since the stability region of the PID parameter space is usually extremely difficult to obtain and be formulated as explicit constraints. Reward modification methods such as Lagrangian relaxation \cite{chow2017risk,bohez2019value} only add some soft penalty on constraint violation or unstable response to discourage such explorations to a certain extent, that is, they cannot strictly ensure that the recommended PID parameters by RL agent for subsequent trials are within the stable region. In this work, we propose a novel idea that is based on a baseline PID controller for ensuring the stability for RL-based PID tuning. 
	
	The proposed method utilizes a conservative but stable PID controller as a \textit{baseline}. The corresponding PID parameters are denoted as $K_{p}^{0}$, $\tau_{I}^{0}$, $\tau_{D}^{0}$, respectively. Conservative controllers enable better tolerance of model uncertainties and disturbances, thereby better capability to maintain the closed-loop stability. The baseline controller can be selected based on prior knowledge of the system and classical tuning guidelines (e.g., Ziegler-Nichols) but with a \textit{de-tuned} setup (thus possibly poor performance). Prior to the training of RL agent, the baseline controller is deployed to the system to observe the closed-loop response over an episode. The tracking error in \eqref{eq: ISE} then can be computed based on the acquired response profiles, and this is used as the \textit{benchmark} reward $R_{bmk}>0$ for the subsequent RL training. The details of the stability-preserving PID tuning are shown in the flowchart in Fig. \ref{fig: Flow_Chart}. As shown in this figure, for the $k$-th RL exploration with PID parameters $K_{p}^{k}$, $\tau_{I}^{k}$, $\tau_{D}^{k}$, initially this combination of parameters is deployed to the controller for closed-loop operation. To ensure stability, a supervisor mechanism is mounted on top of the closed-loop system to monitor the process variables at each operation step. This supervisor accumulates the running reward (RR) or running tracking error at each operation step and compares it with the benchmark reward $R_{bmk}$. For instance, the RR at time $t$ can be defined as (note about the sign):
	\begin{equation}
		RR(t) = \sum_{s=0}^{t} e_s^{2}. \label{eq: RR}
	\end{equation}
	At time $t$, if the running reward exceeds the benchmark reward: $RR(t) >\lambda R_{bmk}$ where $\lambda \ge 1$, it indicates that the underlying PID controller gives a worse performance even than the baseline controller. This occurs with a high chance due to an unstable controller. Upon such observations, the supervisor will replace the underlying parameters by the baseline parameters $K_{p}^{0}$, $\tau_{I}^{0}$, $\tau_{D}^{0}$ to correct the closed-loop response in advance before actual instability happens. The remaining closed-loop operation will be under the baseline controller. Due to the transients afterwards, the eventual total reward will be a much worse value than the benchmark. This final reward will be used as the associated reward for the deployed PID parameters. The obtained poor reward value can inform the RL agent of avoiding exploring nearby regions for subsequent trials, so as to discourage exploring unstable regions. In the subsequent sections, the deployed PID parameters at the beginning of the episode is defined as \textit{explored parameters}, whereas the actually utilized parameters until the end of the episode is defined as \textit{implemented parameters}. For an unstable PID controller, the implemented parameters are indeed the same as the baseline parameters.  The conservativeness of the baseline controller can, to the maximum extent, prevent the process variables from divergence and drive them to the steady-state. The developed stability-preserving DPG-based PID tuning algorithm is shown in Table \ref{Algo_DPG_PID_Tuning}. The theoretical proof of the closed-loop stability of our method can be found in the Appendix.

	\textbf{Remark 1.} Similar to the definition of reward \eqref{eq: ISE}, the running reward defined in \eqref{eq: RR} can be flexible by including other performance metrics, e.g., penalty on the inputs and even the input change rate. However, the form of the running reward shall be consistent with that of the reward in \eqref{eq: ISE}. The underlying principle of using the performance metric as the reward for the RL agent to learn stays the same. The selection of $\lambda$ can be properly determined based on the conservativeness of the baseline controller or other piratical considerations. 
	
	\begin{table}[tb]
		\centering
		\caption{Pseudo code for stability-preserving DPG-based PID tuning} \label{Algo_DPG_PID_Tuning}
		\begin{tabular}{ll}
			\hline
			\multicolumn{2}{l}{\textbf{Algorithm: Stability-preserving DPG-based PID Tuning}} \\ \hline
			1: & Input: initial policy parameter $\theta$, Q-function parameter $w$, and empty replay buffer $\mathcal{D}$ \\
			2: & Initialize target networks $\theta^{\prime}\leftarrow\theta$, $w^{\prime}\leftarrow w$. Choose the baseline controller ($K_{p}^{0}$, $\tau_{I}^{0}$, $\tau_{D}^{0}$)\\
			3:&\textbf{Repeat} \\
			4:& Set running reward $RR = 0$. Observe state $s$ and select action $a=clip(\mu_{\theta}(s)+\epsilon, a_{low}, a_{high})$ \\
			& where $\epsilon$ is random noise. The action is the PID parameters $a=(K_{p},\tau_{I},\tau_{D})$ \\
			5:& Perform closed-loop operation under $(K_{p},\tau_{I},\tau_{D})$ \\
			& for $t=0,1,\ldots,T$:  \\
			&\qquad  \textbf{If} $RR(t)>\lambda R_{bmk}$:\\
			&\qquad \qquad  PID $\leftarrow$ ($K_{p}^{0}$, $\tau_{I}^{0}$, $\tau_{D}^{0}$) \\
			&\qquad  One-step simulation forward, update $RR(t)$, append CV, MV, and setpoint values \\
			6:&Observe the next state $s^{\prime}$ and compute the episodic reward $r$ using the closed-loop data \\
			7:& Store the transition $\left(s,a,r,s^{\prime}\right)$ into the replay buffer $\mathcal{D}$ \\
			8: & \textbf{Update RL agent parameters:} \\
			9:&\qquad Randomly select a batch of transitions $B=\left\{\left(s,a,r,s^{\prime}\right)\right\}$ from $\mathcal{D}$ with \\
			& \qquad the number of transitions as $|B|$ \\
			10: &\qquad Compute the target for each sample in $B$: $y=r + Q(s^{\prime},\mu_{\theta^{\prime}}(s^{\prime}),w^{\prime})\in\mathbb{R}^{|B|}$ \\
			11:&\qquad Update the Q-function by one step of gradient descent using cost function w.r.t. $w$: \\
			& \qquad \qquad \qquad $\frac{1}{|B|} \sum_{\left(s,a,r,s^{\prime}\right)\in B}\left(Q(s,a,w) - y \right)^2$ \\
			12:&\qquad Update the policy by one step of gradient descent using cost function w.r.t. $\theta$: \\
			& \qquad \qquad \qquad $\frac{1}{|B|} \sum_{s\in B}Q(s,\mu_{\theta}(s))$ \\
			13:&\qquad Update the parameters $w^{\prime}$ and $\theta^{\prime}$ in the target networks ($0<\rho<1$ but close to 1): \\
			& \qquad \qquad \qquad $w^{\prime}\leftarrow \rho w^{\prime} + (1-\rho) w$, ~~~$\theta^{\prime}\leftarrow \rho \theta^{\prime} + (1-\rho) \theta$ \\
			14:& \textbf{Until Convergence then End Repeat}  \\
			\hline
		\end{tabular}%
	\end{table}
	
	\subsection{Connection to Gain Scheduling}
	RL-based PID auto-tuning as proposed in this work inherits similarities with gain scheduling \cite{stewart2012pragmatic}. Both strategies seek an ultimate goal of achieving online adaptivity of controllers to varying operating conditions. However, they also have important differences. For gain scheduling, a set of scheduling parameters that indicate current operating condition must be measured online. The deployed controller under current scheduling parameters is updated in real-time by, for example, linearly combining baseline controller parameters that were designed beforehand at prescribed operating conditions. For the RL-based PID tuning, the optimal PID parameters are obtained through the online interactions between RL agent and closed-loop system, and thus it does not need to measure any scheduling parameter. In the literature, two strategies have been reported: PID parameters can be updated in real-time at every time step \cite{lawrence2020optimal} or in an episodic mode as in our work. For the former approach, since the PID parameters are changed instantaneously instead of waiting the closed-loop response to evolve for a duration, it is hard to define a reasonable metric to assess the control performance (e.g., overshoot, integrated squared error, etc.) under given parameters. Moreover, the overly fast abrupt switching of PID parameters during RL search may lead to closed-loop instability \cite{lawrence2020optimal}. For our episodic approach, these two issues can be well addressed and one potential issue is the large number of time steps required to complete the training. In fact, one of our future research topics is exactly on how to accelerate the training of RL agent using sample efficient RL techniques to significantly reduce the number of data samples required for the training. In this way, this potential issue can be mitigated.

	\textbf{Remark 2.} Despite the stability-preserving feature of the proposed approach, some potential limitations may exist for its practical applications. One of the main difficulties, as mentioned above, is the relatively low sample efficiency of not only our method but also general RL approaches. We will tackle this problem in our future work using employing sample efficient RL methods to accelerate the training by reducing the number of required samples. Another potential limitation is the determination of hyper-parameters for the proposed algorithm (e.g., learning rate, actor/critic network structures, etc.) that could affect the convergence and sample efficiency. One solution could be to employ existing well-developed RL models as a starting point (similar to the idea of transfer learning) to facilitate the selection of hyper-parameters.

	\begin{figure} [tb]
		\centering
		\includegraphics[width=5.5in]{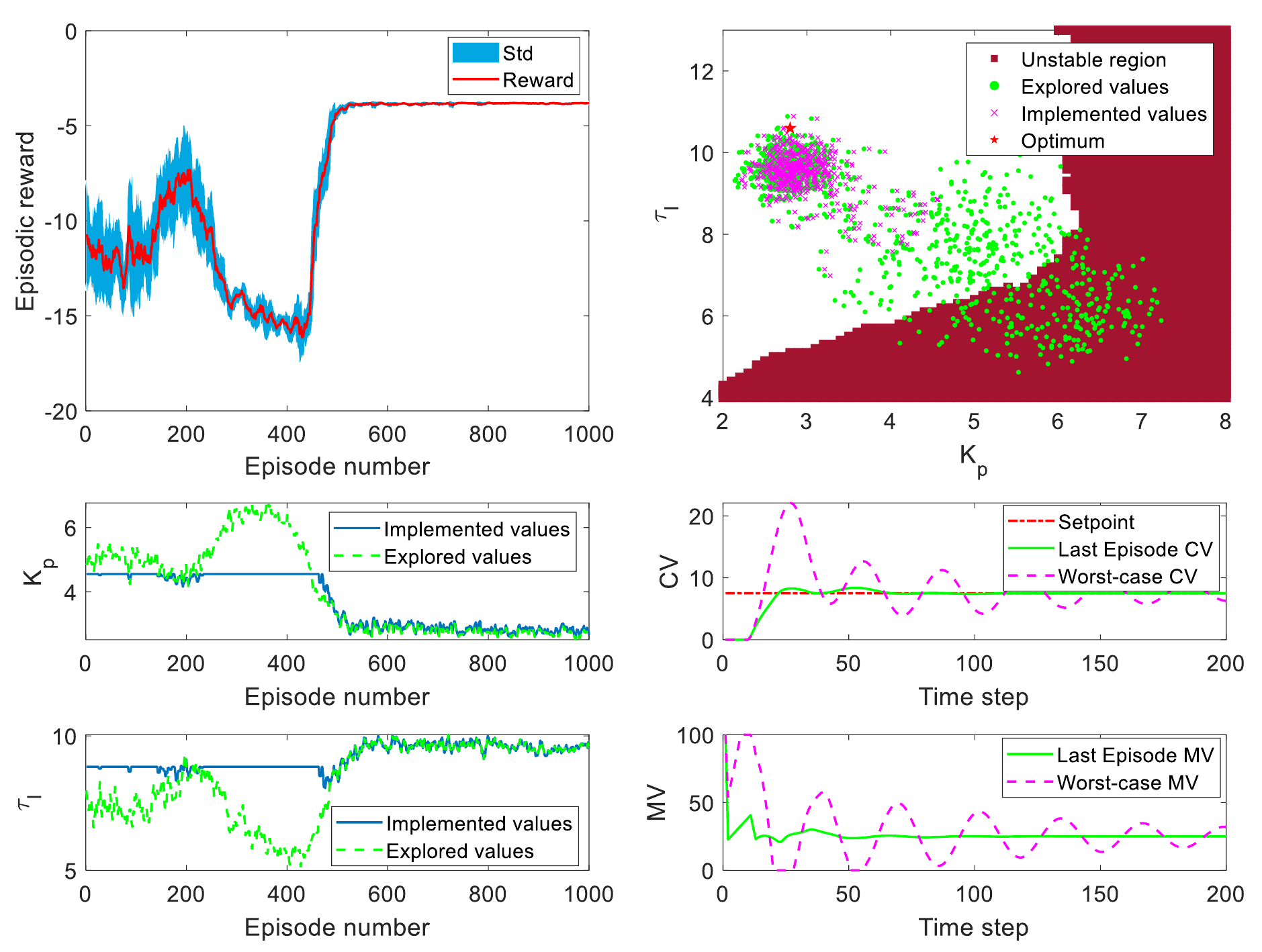}
		\caption{Stability-preserving automatic PID tuning results with Algorithm \ref{Algo_DPG_PID_Tuning} with two tunable parameters $K_p$ and $\tau_I$.}
		\label{fig: Example_2_1}
	\end{figure}
	
	\section{4. Simulation Example} \label{Sec: IV}
	
	In this section, we study the PID tuning for a second-order plus dead-time (SOPDT) process with model as
	\begin{equation*}
		G(s) = \frac{0.3}{25s^2+10s+1}e^{-10s}, 
	\end{equation*}
	with time-delay as 10 seconds. For simulating this process, the sampling time is selected as 1 second. Due to the large time-delay, PID control is the preferred mode. The ranges for the three parameters are respectively $K_p\in [0,10]$, $\tau_I\in [0,15]$, $\tau_D\in [0,10]$. The bound for the MV is chosen to be $u\in [0,100]$. If the computed MV is outside these bounds, the anti-windup technique will be used (cf. Section 2). The episodic length for closed-loop operation (step test) is $T=200$ steps. We use the setpoint-tracking performance as the reward for a given setpoint $r^{sp}=7.5$. The baseline controller is chosen as $K_{p}^{0}=4.56$, $\tau_{I}^{0}=8.85$, $\tau_{D}^{0}=5.90$. In the following case studies, we use numerical methods to separate the stable and unstable regions in the PID parameter space. Specifically, we grid the range of each PID parameter with an interval of 0.2. For each mesh grid of the PID parameter space, we conduct a step test with reference value 7.5 and compute the tracking error. If the tracking error exceeds a threshold (selected as 15 in our work based on our analysis of the tracking errors over several marginally stable PID controllers) and at the same time there is an overshoot, then we claim that the underlying PID parameter combination is unstable. Note that we use overshoot to distinguish the large tracking errors caused by sluggish and unstable controllers. Such an analysis is conducted over all grid points and this gives us the stable and unstable regions. 
	
	\begin{table}[h]
		\centering
		\caption{List of key hyper-parameters used by our RL-based PID tuning algorithm}
		\label{table: Hyperparameters}
		\begin{tabular}{ll|ll}
			\hline
			Parameters & Values & Parameters &  Values \\ \hline
			Learning rate $\alpha$ for actors & 0.0002 & Learning rate $\beta$ for critics & 0.0002 \\
			Memory size $N$ for replay buffer & 1000 & Batch size $B$ & 32 \\
			Episode length $T$ &  200  &  Environment state dimension $n_s$ & 30 \\
			Discount factor $\gamma$ in the return & 0.99  & Polyak averaging coefficient $\rho$ for DPG & 0.999 \\
			Structure for actor and its target networks & [40,30,1] & Structure for critic and its target networks  &  [40,30,1] \\
			Reward threshold $R_{bmk}$ to baseline controller & 15 & Scaling parameter $\lambda$ for reward threshold & 1 \\
			Initial noise variance $\sigma^2$ for exploration & 0.05 & Noise decay factor & 0.001 \\ \hline 
		\end{tabular}%
	\end{table}
	
Table \ref{table: Hyperparameters} lists the key parameters adopted by our algorithm for the following simulations. In determining some parameters such as learning rates, noise level, replay buffer, and batch size, we referred to benchmark DPG models \cite{silver2014deterministic} and did slight tuning on top of those benchmark values. Other parameters such as actor and critic network structure and noise level are determined based on our insight into the specific problem along with some trial and error. Through this process, we discover that the algorithm outcome is sensitive some parameters such as the learning rate and network structure, but not sensitive to parameters such as the buffer size, batch size, etc. Our interpretation is that those sensitive parameters are directly related to the overall complexity, scale, and optimization performance of the resultant RL problem. Therefore, one potential challenge in implementing such RL-based PID auto-tuning methods is the determination of such parameters for the specific application. However to accelerate and automate the selection of hyper-parameters will be an interesting and important future topic.
	
	\textbf{Case 1:} For the first case, PID control is used but only two free parameters $K_p$ and $\tau_I$ are available for tuning. The derivative parameter $\tau_D$ is chosen to be dependent on $\tau_I$, i.e., $\tau_D = 2\tau_I/3$. In other words, the space of tunable parameters only consists of $K_p$ and $\tau_I$. With the developed DPG-based safety-preserving PID tuning Algorithm \ref{Algo_DPG_PID_Tuning}, the simulation results are shown in Fig. \ref{fig: Example_2_1}. The top left plot shows the learning curve of the RL agent during training, where the red curve shows the moving-averaged reward value over episodes and the blue noisy curve shows the raw reward values. It shows that with the proposed method, after initial exploration, the RL agent is able to quickly concentrate on exploitation to converge to the optimum. The top right plot shows the unstable (red) and stable (white) regions of the 2D parameter space. The green dots are the explored parameters by the RL agent during the training, whereas the purple dots show the distribution of the actually implemented parameters. One can see that all implemented parameters are located within the stable region, despite that some explored parameters are in the unstable region. This observation clearly shows the effectiveness of our proposed technique in that it can force the implemented parameters to be always in the stable region to main closed-loop stability. The red star shows the true optimum, which is surrounded by purple dots. This indicates that our algorithm is able to find the global optimum during training. The bottom left plot shows the trajectory of explored and implemented $K_p$ and $\tau_I$ over episodes. At the beginning, the agent explores many unstable parameters but our algorithm can always replace them with the baseline stable controller. The bottom right plot shows the tracking performance under the worst-case episode and the last episode during RL training. It is clear that the PID parameters from the last episode can give a satisfactory tracking performance. For the worst-case episode, due to the  presence of early correction from our stability-preserving algorithm, the CV and MV can still be maintained to be convergent. This observation certifies the effectiveness of the proposed method in preserving closed-loop stability during RL exploration. 
	
	\begin{figure} [tb]
		\centering
		\includegraphics[width=5.5in]{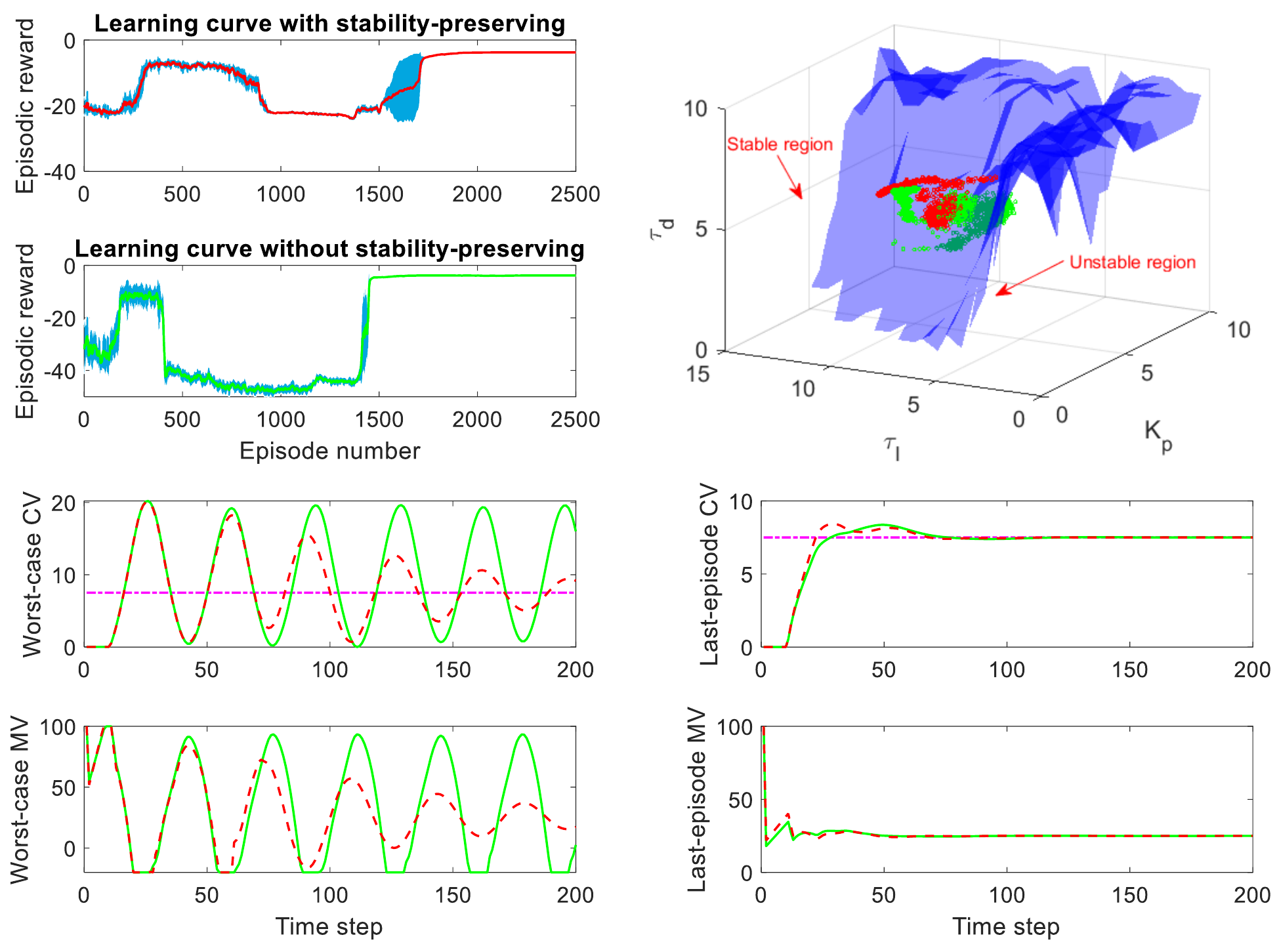}
		\caption{PID tuning results with Algorithm \ref{Algo_DPG_PID_Tuning} with three tunable parameters $K_p$, $\tau_I$, and $\tau_D$. Red: results from stability-preserving RL-based PID tuning; Green: results from traditional RL-based PID tuning without stability-preserving.}
		\label{fig: Example_2_2}
	\end{figure}
	
	\textbf{Case 2:} For the second case, we consider the full-scale setup of PID tuning with three tunable parameters. We use the same baseline controller as the previous case. However, to better clarify the difference between stability-preserving and traditional RL-based PID tuning without stability preservation, in this example, we choose the bounds of MV to be $[-20,100]$ to allow unstable behavior to evolve to certain extent. Note that the presence of MV bounds can make the CV diverge to an upper bound instead of infinity for the unstable response. Fig. \ref{fig: Example_2_2} illustrates the test results with our proposed stability-preserving method (red curves) and traditional method without preserving stability (green curves). The top left plot shows the learning curves (i.e., episodic reward) under these two scenarios. It can be seen that within 2000 episodes, both learning curves converge to a stable value. For the case without stability-preserving, the rewards can be as low as -40 over quite a few number of episodes, whereas for our method with stability-preserving, the lowest reward never exceeds -20. This indicates that our method can avoid exploring some poor regions (possibly unstable regions) during RL search. The top right plot shows the scatter of implemented PID parameter values under these two methods. The light blue surface is the separation between stable and unstable regions in the parameters space, where the volume behind the surface is the unstable region. As shown in this plot, all red points, i.e., the implemented PID parameters with our method, are strictly located in the stable region, whereas some green points (implemented parameters without stability preservation) are behind the surface in the unstable region. The bottom left plot compares the CV and MV of the respective worst-case episode of these two methods during the RL training. It is clear that for our method, even for the worst-case episode, the CV (red) still converges to a steady-state, thanks to the early correction mechanism in our algorithm. In contrast, for the traditional method without considering stability-preserving, the worst-case CV (green) demonstrates oscillation behaviors, which is a clear sign of unstable response. Note that this behavior is indeed strictly unstable since the presence of MV bound prevents the CV to diverge exponentially. The bottom right plot presents the CV and MV of the last episode. Both methods are able to discover a combination of PID parameters with high control performance. As note in \cite{lawrence2020reinforcement}, the MV bound can ensure bounded unstable response so as to give a meaningful reward value to RL agent. Without such a bound, the presence of a portion of divergent responses will significantly deteriorate the data quality in the replay buffer, which will disable the RL agent to learn a useful policy. These results again clearly verify the effectiveness of our stability-preserving RL-based PID automatic tuning framework. 
	
	\begin{figure} [tb]
		\centering
		\includegraphics[width=5in]{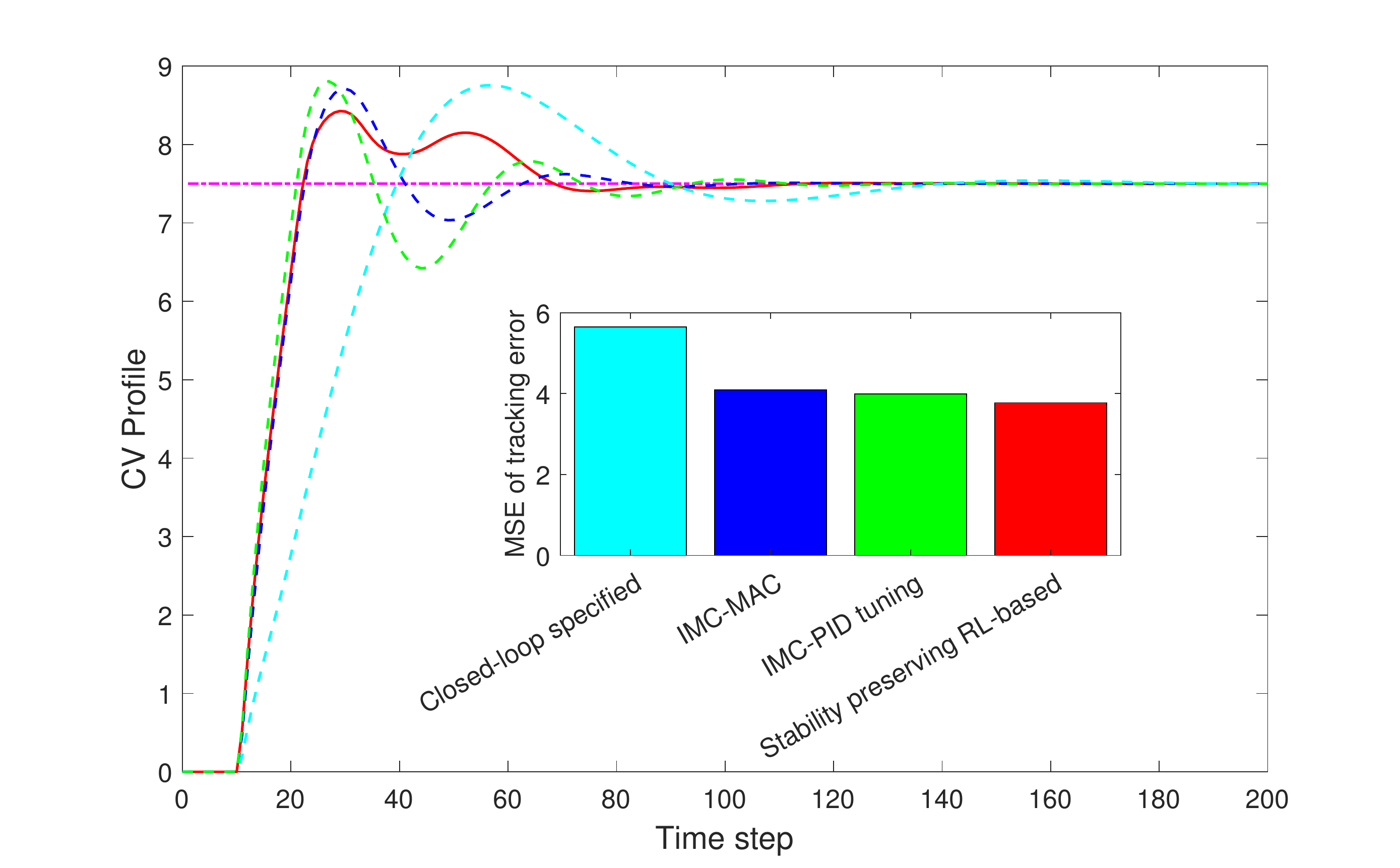}
		\caption{Comparison of the  of RL-based PID auto-tuning (our approach), closed-loop specified, IMC-MAC, IMC-PID tuning method}
		\label{fig: tuning_benchmark}
	\end{figure}

   \textbf{Comparison with existing PID tuning methods}. There are numerous methods available in the literature for PID tuning such as Ziegler-Nichols method \cite{ellis2012four}, direct synthesis method, internal model control (IMC) tuning, IMC-Maclaurin closed-loop tuning \cite{panda2004pid} and so on. In this work, we specifically select the following three classical methods for comparison with our method: IMC-PID tuning, IMC-Maclaurin tuning, and closed-loop specified tuning. Prior to using the IMC-PID tuning method, a first-order plus dead-time (FOPDT) model has to be developed to approximate the SOPDT model. To this end, the following relations are used to obtain the FOPDT approximation model \cite{panda2004pid}:
	\begin{equation}
			K_m=K_c, \quad \tau_m=1.641\tau, \quad D_m=0.505\tau+D, 
	\end{equation}  
	where $K_m$, $D_m$, and $\tau_m$ are the gain, dead time, and time constant of the approximated FOPDT model, respectively. $K_c$, $D$, and $\tau$ denote the gain, dead time, and time constant of the original SOPDT system. The tuning parameters are then obtained by the first row of Table \ref{table: TuningMethod}, where $\lambda = \max(0.25D_m,0.2\tau_m)$ is a hyper-parameter. IMC-MAC and closed-loop specified tuning methods can handle the SOPDT system directly.  The specific formula for determining the PID parameters can be seen in Table \ref{table: TuningMethod}. The tuning parameters calculated from the IMC-PID, IMC-MAC and closed-loop specified PID tuning methods are $[3.08,14.47,3.55]$, $[2.87,12.92,3.99]$, and $[0.833,5,5]$, respectively.

	Fig. \ref{fig: tuning_benchmark} compares the setpoint tracking performance between our method and the other three selected tuning guidelines. Among these step response curves of the output, our method tracks the setpoint quickly despite the presence of a secondary peak. The closed-loop specified tuning gives the worst performance. The embedded bar plot shows that our method yields the minimum tracking error among others. The reason is that our RL-based PID tuning aims at finding the optimum in the PID parameter space to maximize the reward. However, the other classical tuning methods are more empirical and their purpose is to quickly provide a good set of PID parameters without explicitly optimizing a specific objective. 
	
	\begin{table}[h]
		\centering
		\caption{The selected classical PID tuning methods for SOPDT model}
		\label{table: TuningMethod}
		\begin{tabular}{ll}
			\hline
			Method & PID Parameter Determination  \\ \hline
			IMC-PID tuning \cite{panda2004pid} & $K_P=\frac{2\tau_m+D_m}{2K_m(\lambda+D_m)},  \tau_I=\tau_m+0.5D_m,  \tau_D=\frac{\tau_m D_m}{2\tau_m+D_m}$  \\
			IMC-MAC tuning \cite{rivera1986internal} & $K_P=\frac{\tau_I}{K_c(2\lambda+D)},  \tau_I=2\tau-\frac{2\lambda^2-D^2}{2(2\lambda+D)},  \tau_D=\tau_I-2\tau+\frac{\tau^2-\frac{D^3}{6(2\lambda+D)}}{\tau_I}$  \\
			Closed-loop specified tuning \cite{lee1998pid} & $K_P=\frac{\tau}{2K_cD}, \quad \tau_{I}=\tau, \quad \tau_D=\tau$  \\
			\hline
		\end{tabular}%
	\end{table}
	
\begin{figure} [tb]
		\centering
		\includegraphics[width=3in]{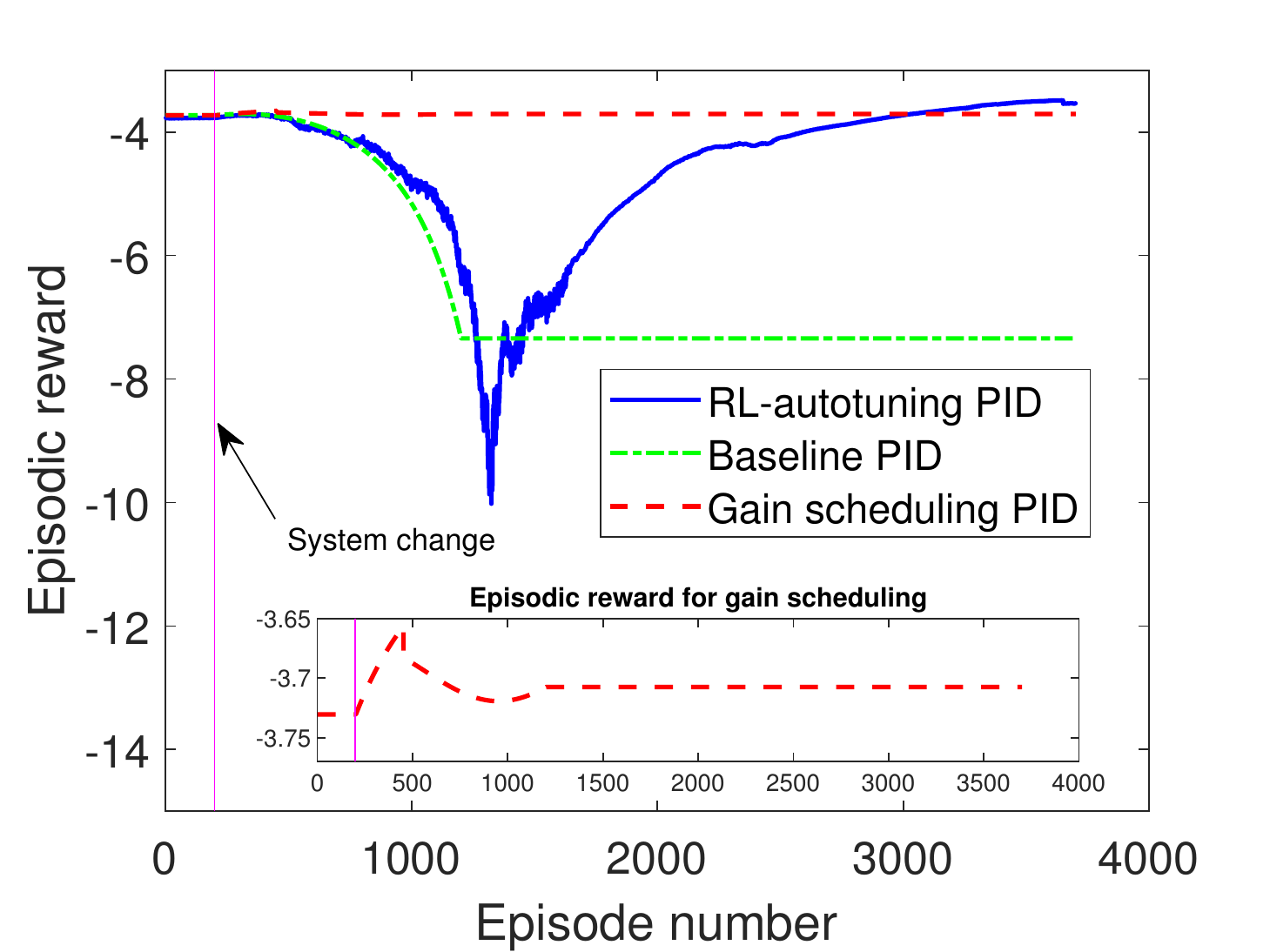}
		\includegraphics[width=3in]{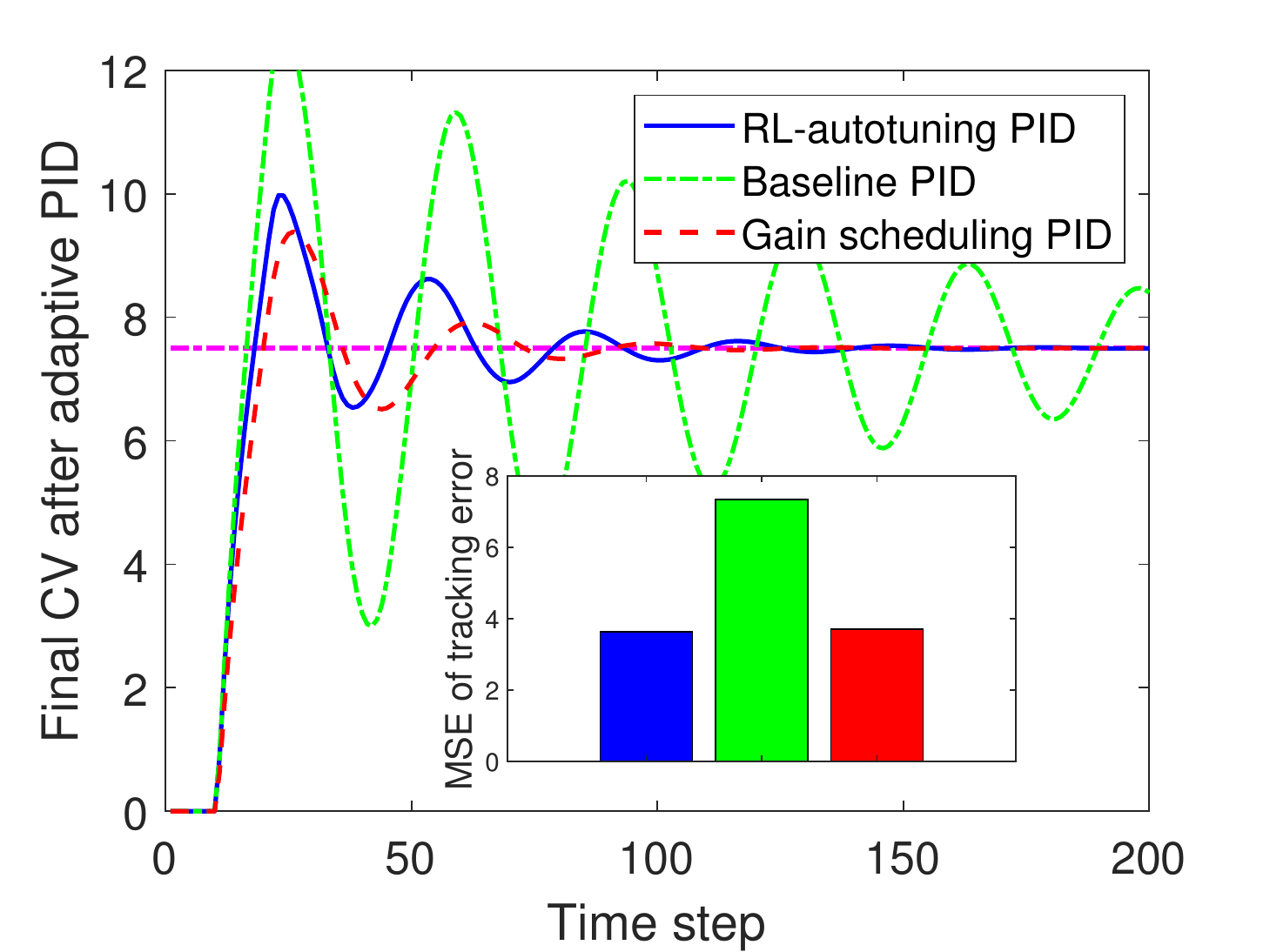}
		\caption{Comparison of the adaptivity of RL-based PID auto-tuning (our approach), gain-scheduling PID, and the fixed baseline PID controller to the process gain change. Left: episodic rewards (the negative of mean squared tracking error) for these methods; Right: The ultimate CV profiles after the PID adapts to the new process gain based on these methods. Note that that green color shows the results where a fixed baseline PID from Case 2 is used throughout the entire experiment.}
		\label{fig: Example_2_3_Comparison}
	\end{figure}

\begin{figure} [tb]
		\centering
		\includegraphics[width=5.5in]{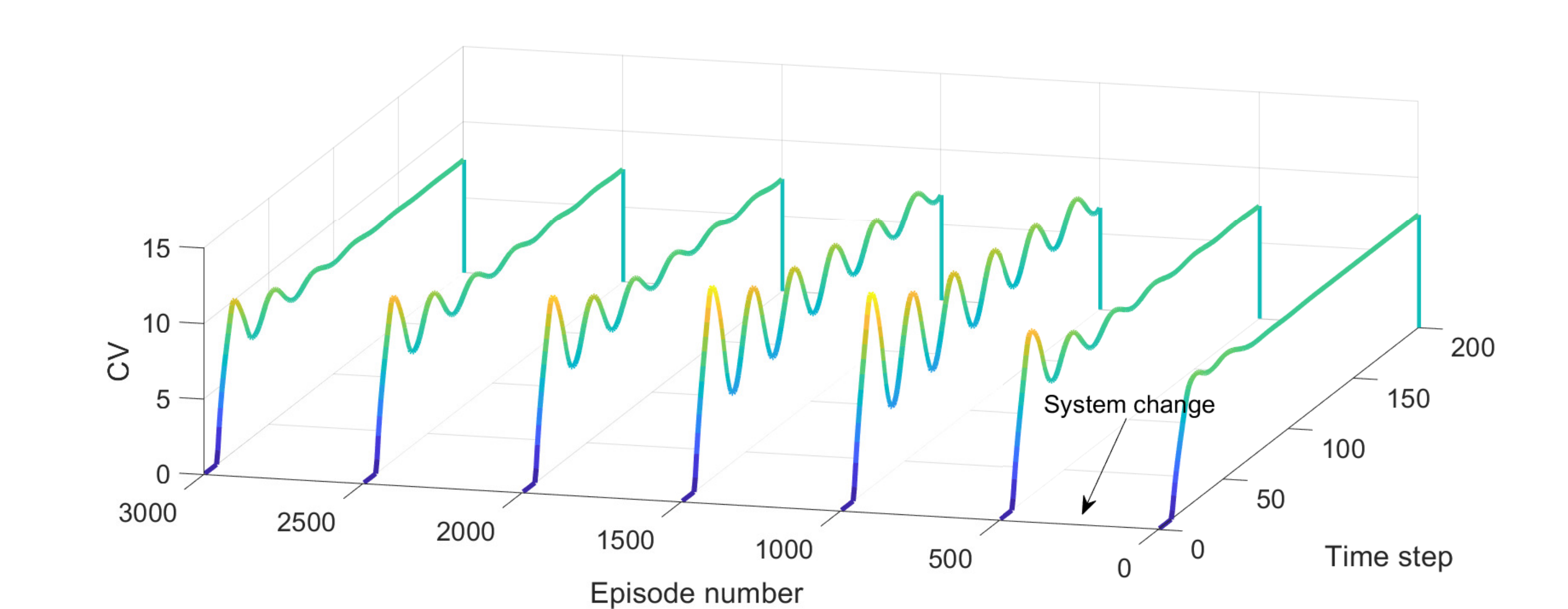}
		\caption{Evolution of the CV profiles with RL-based PID auto-tuning after introducing the process gain change.}
		\label{fig: Example_2_2_Evolving}
	\end{figure}

	\textbf{Adaptivity to changes in system model}. In this subsection, we extend the previous results from Case 2 to study the adaptivity of our stability-preserving RL-based PID tuning. To this end, after the initial RL training, we purposely introduce a change in the system gain. We then examine whether the RL agent is able to adapt to the new system via interactions with the closed-loop system. Specially, this simulation experiment starts after the episodic reward in the top left plot of Fig. \ref{fig: Example_2_2} settles down, e.g., after 2000 episodes. With the trained RL agent, we operate the closed-loop system in an episodic mode and gradually change the process gain from 0.3 to 0.5 in 1000 episodes. Such a slow transition in the system parameter is analogous to practical situations when the process operation condition slowly drifts over time. If a scheduling parameter is measurable (e.g., if the process gain is measurable for our case) to indicate the current operating condition in real-time, gain scheduling control will be a variable approach to enable the adaptivity of controllers \cite{stewart2012pragmatic}. In this subsection, we compare our RL-based PID auto-tuning approach with gain-scheduling PID by assuming that the process gain (i.e., the scheduling parameter) is measurable in real-time. Note that our RL-based PID tuning does not require any measurable scheduling parameters, in contrast to gain scheduling. For the gain-scheduling PID control, several benchmark controllers under prescribed operating conditions shall be designed in advance. Then, the real-time PID controllers can be obtained by linearly interpolating those baseline controllers based on the current value of the scheduling parameter. For simplicity, we pre-designed two PID controllers under two prescribed conditions where the process gains are 0.3 and 0.6, respectively. We used the guideline provided by \cite{seborg2016process} for SOPDT to design two benchmark PID controllers as $[2.87,12.92,3.99]$ and $[1.44,12.92,3.99]$,  respectively. The real-time PID controller parameters are obtained by interpolating these two benchmark controllers based on the current process gain value. The left plot in Fig. \ref{fig: Example_2_3_Comparison} shows the control performance (the reward is the negative of the mean-squared tracking error) under different adaptive PID strategies. For our approach shown in blue, the reward initially drops after the change of the process gain, however, through a number of interactions, the RL agent is able to learn the best PID parameters and recovers the control performance under the new operating condition. The gain-scheduling PID presents consistent control performance in the presence of process gain change (the embedded subplot is a microscopic view), which gives the best performance among the three methods. However, this observation is based on the assumption that we have a measurable scheduling parameter in real-time, which does not always hold in practical situations. As another comparative method, we consider using the previous baseline controller from Case 2 with fixed PID parameters throughout the entire experiment. As shown in green, this fixed controller is not able to adapt to the system change, thus giving a degraded control performance which never recovers. The right plot of Fig. \ref{fig: Example_2_3_Comparison} presents the final CV profiles from these adaptive PID strategies after adaptation. Our approach even gives a slightly better control performance than the gain-scheduling PID, as indicated by the bar plot. The fixed baseline PID controller presents the worst performance. To visualize the adaptation process of the RL-based PID auto-tuning, we plot the evolution of the CV profiles before and after introducing the process gain change, as shown in Fig. \ref{fig: Example_2_2_Evolving}. From this figure, after the system change, the control performance starts to deteriorate, since the previous well-tuned PID parameters may not perform well under the new operating condition. The RL starts to gain experience from the degraded rewards, after a number of interactions, eventually, the RL agent is able to generate good PID parameters to adapt to the new operating condition and recover the control performance (see episode \# 3000).
	
	\textbf{Remark 3.} Similar to other RL-based methods as well as shown by the results from this simulation section, the major computation burden of the proposed algorithm lies in the training of the RL agent. For a large and complex PID parameter space, the exploration of the space and discovering the optimum can take a large number of training samples. Indeed, there are a number of factors that can directly affect the training time, such as the learning rate, episode length, and action noise level. If the learning rate is overly small together with a long episode, the RL agent would take more time to obtain the state and reward from the closed-loop system and require more iterations to converge to the optimum, thus can considerable increase the training time. In addition, if the action noise level is too large, then RL's exploration of the PID space can consume much more  time. Therefore, a good trade-off between exploration and exploitation is also an important factor that affects the computation complexity of the proposed method.
	
	\textbf{Remark 4.} In this section, we used a linear second-order system as an example to implement and validate our algorithms. Although the selected system is simple, our main objective here is to verify that the proposed stability-preserving approach can ensure closed-loop stability during the search of optimal PID parameters, as indicated in the results above. However, our ultimate long-term objective is to achieve a fully online automatic PID tuning framework for generic nonlinear systems (e.g., CSTR) based on RL methods to enable the online adaptivity, optimality, and stability throughout varying operating conditions. As part of our future work, the proposed stability-preserving method will be applied to nonlinear systems where online auto-tuning will also be achieved. The determination of closed-loop stability can be conducted via monitoring the running reward, instead of relying on theoretical approaches that can be nontrivial for nonlinear systems.
	
	\section{5. Conclusions} \label{Sec: V}
	We presented a novel stability-preserving framework for RL-based automatic PID tuning. This work is motivated by the observation that although methods exist for the PID tuning for complex systems, many conventional PID tuning methods rely on trial and error. The obtained PID parameters from such empirical methods may not be optimal. Moreover, existing RL-based tuning methods cannot ensure the closed-loop stability during policy search. To this end, we propose a novel multi-scale episodic PID automatic tuning framework based on the DPG algorithm. For this method, the RL agent receives the reward and updates the network parameters once after each entire closed-loop operation. To preserve the closed-loop stability, we employ a stable PID controller as a baseline whose reward is used as a benchmark. A supervisor mechanism monitors the running reward for any explored PID parameter. Once the running reward exceeds the benchmark, the supervisor mechanism replaces the underlying PID controller with the baseline controller to prevent unstable response. Simulation examples are provided to show that our method can not only discover the optimal tuning parameters efficiently but also preserve the closed-loop stability for all explored PID parameters, compared with the standard RL-based PID tuning without considering stability preservation. Moreover, the proposed method is able to adapt to system changes without requiring any knowledge about the underlying operating condition. Future work includes enabling the online and efficient PID tuning with RL-based methods for nonlinear systems while ensuring stability during the policy search.
	
\section{Appendix}\label{sec: Appendix}
This appendix is devoted to the proof of the overall closed-loop stability of our stability-preserving method.
	
We consider a generic nonlinear system 
\begin{equation}
			\dot{y} = f(y(t),u(t)) ,
\end{equation}
where $y(t)$ and $u(t)$ are, respectively, the CV and MV at time $t$. Assume that the initial unstable controller delivered from RL is expressed as $u(t)=k_u(y(t))$, and the baseline stable PID controller is $u(t)=k_s(y(t))$.  Then the closed-loop system can be formulated as 
\begin{align}
			\dot{y} &= \begin{cases}
				f(y(t),k_u(y(t)))=g_u(y(t)),& \text{if} \quad t < t_1 \\
				f(y(t),k_s(y(t)))=g_s(y(t)),& \text{if}\quad t \ge t_1,
			\end{cases}
			\label{eq: nonlinear}
\end{align}
where $t_1$ the instant when the running reward exceeds the threshold for the first time. To prove the overall closed-loop stability with a switching from unstable to stable PID controller, we assume that the baseline PID controller is exponentially stable, which is introduced as below. 
	
\textbf{Definition (Exponential stability)}. \cite{khalil2002nonlinear} Consider a generic autonomous nonlinear closed-loop system $\dot{y} = g(y(t))$. The equilibrium $y=0$ (i.e., g(0)=0) is exponentially stable if there exist constants $m>0$, $\alpha>0$, and $\varepsilon>0$ such that
\begin{equation}
			\|y(t)\| \le m e^{-\alpha(t-t_0)}\|y(t_{0})\|, \label{eq: stability_definition}
\end{equation}
for all $\|y(t_0)\| \le \varepsilon$ and $t\ge t_0$. 
	
For the proposed stability-preserving strategy, define the reward threshold for switching into the baseline controller as $\delta>0$. Due to the finite threshold $\delta$, the output at time $t_1$ is bounded, i.e., $y_1=y(t_1)$, $0<\delta \le \|y_1\| \le \bar{y}$.  Notice that for any $t>t_1$, the stable PID controller is employed, and thus the definition of exponential stability applies: there exist constants $m>0$ and $\alpha>0$ such that
\begin{equation}
			\|y(t)\| \le m e^{-\alpha(t-t_1)}\|y_1\|\le m\bar{y} e^{-\alpha(t-t_1)},~~\forall t>t_1. \label{eq: bound}
\end{equation}
Note that the switching time $t_1$ differs depending on the initial value $y(t_0)$ and the specific unstable controller $u(t)=k_u(y(t))$ delivered from the RL. However, as long as the initial value $\|y(t_0)\|<\delta$, the stable baseline controller will replace the unstable controller and thus maintain the above exponential bound to be valid. This exponential bound converges to zero over time and this ensures the overall closed-loop exponential stability. 
	
	\section{Declarations}
	
	\subsection{Acknowledgments}
	
	M.A. Chowdhury acknowledges the support of Distinguished Graduate Student Assistantships (DGSA) from Texas Tech University. Q. Lu acknowledges the new faculty startup funds from Texas Tech University. All the authors acknowledge the valuable comments and suggestions from anonymous reviewers.

	\subsection{Authors' contributions}
	
	
	Concept development and data acquisition: A.I. Lakhani, Q. Lu \\
	Drafting the manuscript: Q. Lu, M.A. Chowdhury \\
	Modifying the manuscript: A.I. Lakhani, M.A. Chowdhury \\
	
	\subsection{Availability of data and materials}
	Data will be made available upon e-mail request to the corresponding author.

	\subsection{Financial support and sponsorship}
	
	
	None.

	\subsection{Conflicts of interest}
	
	All authors declared that there are no conflicts of interest.

	\subsection{Ethical approval and consent to participate}
	
	Not applicable

	
	

	\subsection{Copyright}
	
	© The Author(s) 2021.
	
	\bibliographystyle{oae}
	\bibliography{scalab}

\end{document}